\documentclass{ws-procs9x6}
\usepackage{psfig}

\begin{document}

\title{Consistency between the radio \& MIR source counts using the radio-MIR correlation}

\author{N. SEYMOUR}

\address{Institut d'Astrophysique de Paris, 98bis, Boulevard Arago, 75014 Paris, France}

\author{I. M$^{\rm c}$HARDY and K.F. GUNN}

\address{School of Physics \& Astronomy, University of Southampton, \\
Highfield, Southampton, SO17 1B, UK}


\maketitle

\abstracts{We show from the recent extrapolation of the 
radio-FIR correlation to the MIR that the 20 cm and 15 $\mu$m differential 
source counts are likely to come from the same parent population.}

\begin{figure}[ht]
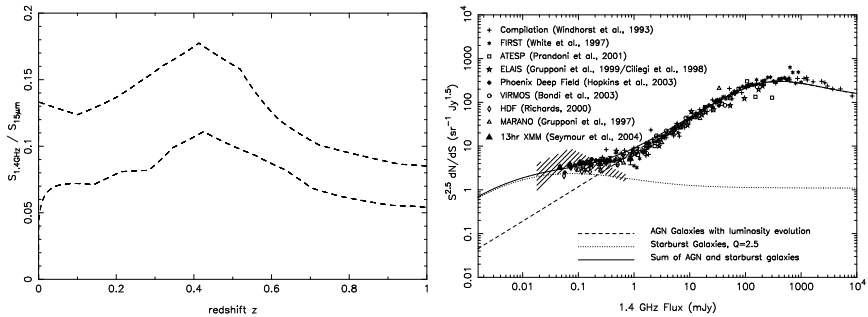

  \begin{center}
    \begin{minipage}[l]{0.49\linewidth}
      \psfig{file=gzd.ps,height=4.1cm,angle=270,width=\linewidth}
    \end{minipage}\hfill
    \begin{minipage}[tl]{0.49\linewidth}
      \psfig{file=final_logr.ps,height=4.1cm,angle=270,width=\linewidth}
    \end{minipage}\hfill
  \end{center}
  \caption{{\bf Left} The 20 cm/15 $\mu$m observed flux density ratio derived 
from the radio/MIR correlation of Ref. 2 as a function 
of redshift (upper and lower limits from error in radio/MIR correlation). 
{\bf Right} The differential 20 cm source counts from the literature with 
the hatched region indicating the superposition of the 15 $\mu$m counts.}
\label{fig}
\end{figure}

\vskip -0.4cm
\noindent
{\bf The Radio-FIR correlation and source counts}
\noindent
The radio-FIR correlation is one of the tightest observational results 
in astronomy\cite{carilliyun00} and has recently been extended to the MIR 
(directly\cite{gruppioni03}, and indirectly\cite{garrett02}). This 
relationship is interpreted as being due to different physical 
manifestations of star-formation at different wavelengths. 
The differential number counts at 20 cm have long been known to have an 
up-turn below 1 mJy\cite{windhorst90}, oft explained by the dominance of 
star-forming galaxies over AGNs (which dominate at higher flux densities). 
The 20 cm counts have been modelled by 
Ref.~\refcite{hopkins} \& Ref.~\refcite{seymour} where 
luminosity-evolution of the local star-forming radio luminosity function 
must be evoked to explain the shape of the curve. The 15 $\mu$m 
differential source counts also show an up-turn below 1 mJy which exceed the 
counts from a non-evolving local luminosity function\cite{elbaz}. This excess 
at 15 $\mu$m is also thought to be due to star-formation.
\newline
\newline
\noindent
{\bf Extrapolation to 20 cm of the 15 $\mu m$ source counts}
\noindent
The fit of Ref.~\refcite{elbaz} to the 15 $\mu$m source counts can be 
superimposed upon the 20 cm source counts, to test for consistency, by 
applying the flux density ratio of objects which follow the radio-MIR 
correlation. To derive this ratio, 20 cm luminosities were K-corrected 
assuming a radio spectral index of $\alpha=-0.7$ ($S\propto\nu^{\alpha}$) 
and 15 $\mu$m luminosities were K-corrected using an M82-like spectrum 
(i.e. star-burst) for the LW3 ISOCAM bandpass\cite{Franceschini}.
The flux density ratio is found to be roughly constant (Fig.~\ref{fig}, left) 
as a function of redshift over the range used by Ref.~\refcite{gruppioni03} 
for the radio-MIR correlation, $z<0.7$. By using a flux density ratio of 0.12 
the contribution of star-forming galaxies observed at 15 $\mu$m is consistent 
with that at 20 cm (Fig.~\ref{fig}, right).

\vskip -0.5cm
\noindent
{\bf Conclusions and Comment} 
\noindent
We have shown that the source counts at 20 cm 
and 15 $\mu$m below 1 mJy and the correlation of their rest frame luminosities 
are all consistent, suggesting that they are from the same parent population, 
possibly dominated by star-formation. We note though that this consistency 
depends 
strongly on the normalisation of the radio-MIR (cf the higher normalisation 
of Ref.~\refcite{garrett02}) and also partly on the SED used for the 15 $\mu$m 
K-correction. Detailed analysis of the faint source population using 
{\it Spitzer} will shed light on the true relative contribution of 
star-formation and AGN within faint objects collectively and individually.

\end{document}